\documentclass[a4paper,12pt]{article}
\usepackage{graphicx,amssymb}
\usepackage[T1]{fontenc}
\usepackage[english]{babel}

\usepackage{hyperref}

                      \def\g{\gamma}
\def\d{\delta}          \def\eps{{\epsilon}}     \def\ve{\varepsilon}
         \def\k{\kappa}           \def\l{\lambda}
\def\m{\mu}

\def\G{\Gamma}                     \def\L{\Lambda}

\def\be{\begin{equation}}            \def\ee{\end{equation}}
\def\ba#1{\begin{array}{#1}}         \def\ea{\end{array}}

\def\fr#1#2{\textstyle\frac{#1}{#2}}

\begin{document}

\renewcommand{\refname}{\bf\large References}
\pagestyle{empty} \noindent
{\Large\bf Absolute parallelism, modified gravity,}\\[1.5mm]
{\Large\bf and suppression of gravitational
{\em short} waves}\\[2.5mm]
{\bf I.L.\,Zhogin} \ (E-mail: zhogin@mail.ru; \
 \href{http://zhogin.narod.ru}{http://zhogin.narod.ru}) \\[1mm]
ISSCM SB RAS, Kutateladze 18, 630128
 Novosibirsk, Russia\\[-1.5mm]

 {\small \noindent
 There is  a unique variant of Absolute Parallelism, which
  is very simple as it has
  no free parameters: nothing (nor $D{=}5$)
 can be changed if to keep the theory safe from emerging
singularities of solutions.
 On the contrary,  eternal solutions of this theory,
 due to the linear instability of the trivial
 solution, should be of great complexity which
    can in some scenarios (with a set of slowly varying
    parameters of solutions) provide
 a few phenomenological models
including a modified (better to say, new or another) gravity and
an expanding-shell cosmology (the longitudinal polarization gives
the anti-Milne model).
 The former looks (mostly) like $R^{\mu\nu}G_{\mu\nu}$-gravity
 on a brane of a huge scale $L$ along the extra-dimension.

 The correction to Newton's law of gravity, which depends
  in this theory  on two parameters (bi-Laplace equation)
 and behaves as $1/r$ on  large scales,
 $r\,{>}\,L$\,(kpc\,${>}\,L\,{>}$\,pc),
  can start from zero (the Rindler term vanishes)
  if a constraint is imposed on these parameters.
On further consideration, one can conclude that generation of
gravitational `short' waves, $\lambda\,{<}\,L$, is inhibited in
this
 new gravity.}\\[-0.5mm]

\noindent {\bf\large  1. Introduction}\\[-2.5mm]

\noindent
 The basement of the modern physics is composed of two main
 theories: the general relativity theory (GR; the leading and
 supposedly successful, at the expense of inventing those dark
 entities, gravitation theory) and the standard model
 of elementary particle physics (a variant of quantum field
 theory which has a great number of unexplained features and
 fitting parameters).
 These two are based on very different principles and
 even symmetries, but they have something in common: both are
 affected by the problem of singularities or divergencies
 (despite different shifts and tricks,
  like supersymmetry, strings, and so on) of
 solutions.
 It is  generally agreed that a more
 fundamental theory should unite these two branches, two domains
 of natural phenomena; and it is deadly evident that a reasonable
 theory should be free from singularities of
 solutions.\footnote{Some gauge dials
 carry the infinity sign $\infty$, but it's just an
 exaggeration -- no one can measure infinity.}

 There are too many interpretations of quantum
 mechanics;\footnote{The most weird one imputes free will
 to elementary particles (= different tiny spaceship  models); see
  arXiv:physics/0004047  by R.\,Nakhmanson; sure, the degree of
  arbitrariness and unknownness in this model is some
  greater than in string theory, or in the strand model
  (motionmountain.net/research.html).}
   two of them -- Copenhagen and many-worlds -- are of
   pretty similar
   `rating' (this means that none is convincing enough).

 Einstein was not satisfied with GR (and quantum
 mechanics as well)
 and he had proposed absolute parallelism (AP) which unites
 symmetries of both general and special relativity theories.
 Einstein and Mayer had obtained a vast list of compatible second order
 equations of (4$D$) AP \cite{eima}, most of them are non-Lagrangian;
 however this list is a bit incomplete.

 Theoretical physicists form in fact a quite specific
subset of
 experimentalists:
 they are doing experiments on their own brains. The mainstream
 theorists participate in highly collective `experiments'.
 However, as one should note,  string theory, as well as M-theory,
  still does not
 deserve the definite article, {\sl the}. (That is, nobody can
 answer, what is string theory?)

 The result of my own `experiment' is a single-field theory,
 really simple (according to Kolmogorov's
theory of algorithm complexity) and
 beautiful (that is, of very large symmetry) --
 the unique (no free parameters)
  5$D$ variant of Absolute Parallelism, which is free
 from emerging singularities in
 solutions of general position \cite{on}.
 (The ``Little Prince's Principle'' states: true beauty should be
single, should be unique. A theory with free parameters, or even a
single parameter, is actually a huge set of slightly different
theories, only one of which is supposedly true, but that true
value(s) will never be measured ``exactly'', at least because of
funding limitations; and there are no reasons to see ``less
beauty'' in all other, wrong variants of such a theory.
Phenomenological models, with a set of `free parameters', of
course, are of some usefulness, but what we really need is a
really fundamental theory which would provide us with all
necessary and well-behaved phenomenologies.)

 Interestingly, in the
absence of singularities,  AP obtains topological features a tad
similar to that of nonlinear sigma-models.
In order to give a clear presentation and full picture of the
theory, many items should be sketched: linear instability of the
trivial solution and expanding O$_4$-symme\-trical solutions;
tensor $T_{\mu\nu}$ (positive energy, but only three polarizations
of 15 carry  $D$-momentum and angular momentum; how to quantize ?)
 and post-Newtonian effects; topolo\-gical
classification of symmetrical 5$D$ field configurations (alighting
on evident pa\-rallels with the particle combinatorics and chiral
patterns of the Standard Model) and a `quantum phenomeno\-logy on
an expanding classical background';
 a `phenomenological' $R_{\mu\nu}G^{\mu\nu}$-gravity on a very
thick brane and a change in the Newton's Law: $\fr1{r^2}$ goes to
$\fr1{r}$ with distance. (This is different from the MOND paradigm
\cite{mond}.  MOND is not free of a strangeness: given two bodies
of very different mass, one can choose the distance between them
such that the heavier body is in the MOND regime, while the other
still in the Newtonian regime -- as a result, the third Newton's
law, if not the second which is testable
 \cite{n2}, should be
violated. Moreover, MOND means violation of linearity exactly for
the case of small accelerations.)

  The linear instability of the
 trivial solution, a very interesting feature, makes inevitable
 the existence of strong non-linear effects; interestingly, the unstable
 waves (growing polarizations) do not contribute to the energy-momentum.
 On the other hand, a spherically-symmetrical single wave running
 along the radius (the longitudinal polarization, by itself, is stable)
  can serve as a region of enhanced instability, a kind of
  wave-guide with a specific (ultrarelativistic) reduction
  of the extra-dimension. A Lagrangian phenomenology
  of topological
  quanta (a kind of topological
  Brownian particles which carry topological charges
  and/or quasi-charges \cite{on,tc})
  emerges, which can look like a Quantum Field Theory
   (although the underlying theory
  is absolutely classical~!).

 The theory seems able to easily explain
  the meaning of many features
 of the Standard Model -- the flavors and
 colours,  quark confinement,  ubiquity of
  the least action principle (and the very superposition
  principle as well). Moreover,
   this theory gives a number of testable predictions:

 \begin{itemize}
 \item  spin zero elementary particles do not exist;

 \item neutrinos are true neutral (a kind of Majorana);

 \item there is no room for SUSY, Dark Matter, and Dark Energy;

 \item additional pseudovector bosons (responsible for
  dynamical mass generation) can exist;

 \item the gravitational part of the Lagrangian is
 $R^{\mu\nu}G_{\mu\nu}$, and, due to the large extra-dimension,
 it gives switching from  Newton's law,  $1/r^2$
 (at small distances), to the more slowly decreasing force,
 $1/r$, at larger distances;

 \item the Hubble plot should be described by the anti-Milne model
 [in FRW-framework it means $a=a_0(1+H_0 t), k=+1$],
 without any fine tuning
 and free parameters (excepting the Hubble constant, of course).

 \end{itemize}

 Frankly, at the present time,
  this theory has seemingly some more reasons, than any other
 existing one, for a belief that it is {\sl on the right track}.
 Here I am going to add one more qualitative prediction
 relating to the subject  of gravitational waves:

 \begin{itemize}
 \item  generation of gravitational `short' waves
 ($\lambda{<}L$) is suppressed.
  \end{itemize}
 But first let's dwell a bit on polarization degrees
 of freedom in GR and
 modified gravities.  \\[-1mm]

 \newpage\noindent
 {\bf\large 2. AP vs GR and Riemann-squared gravities;
 polarization degrees } \\[-2.5mm]

  \noindent
 The case of vacuum GR  is a simple example of a single-field
 theory -- the components of the metric field transform
 as a single irreducible representation;
  also it is a  special, degenerate case of AP.
 Symmetries of the vacuum equation of GR, besides the local group
 of coordinate diffeomorphisms, Diff$(D)$, include the global
 symmetry --  global conform transformations.

 If $g(x)$ is a solution, than $\k^2 g(x)$
  (where $\k{=}$const) is
 also a solution. This simple  symmetry leads to (or explains)
   the notion of [length]
 dimension  which is really  a global
 feature.\footnote{Length is so in Africa as well (i.e.,
 dlina -- ona i v Afrike dlina).}
 Accompanying this, $g_{\mu\nu} \to \k^2 g_{\mu\nu}$,
   with the scale change of coordinates,
 $x^\mu \to \k x^\mu$, one easily obtains
 \be \label{glob}
 g_{\mu\nu} \to \k^0 g_{\mu\nu}; \ g^{\mu\nu} \to \k^0 g^{\mu\nu};\
 g_{\mu\nu,\lambda} \to \k^{-1} g_{\mu\nu,\lambda};
 \mbox{ \ et cet.}
 \ee
 The power of factor $\k$ corresponds to the power of {\em
 length} indicating the dimension of one or another value.
 The moral is that  global symmetries could also be very important
 (and  intention to introduce the local conform symmetry is not so
 reasonable).
 In the presence of the cosmological constant, the GR equation is
 no longer homogeneous, and the global conform symmetry
 disappears.

 There is, however, another noteworthy global feature in nature
 --- the signature of spacetime. In special relativity, it follows
 from the global symmetry, Lorentz group, but it has no such a
 `symmetry
 substantiation' in the case of general relativity. The absolute
 parallelism theory (AP) improves the situation.
  The frame field of AP, $h^a_{\,\mu}$,
 is just a square matrix with indexes of different nature; it
  admits both local coordinate transformations (act on Greek indices)
   and global
   transformations of `extended' Lorentz group (the
   point symmetry group of inertial coordinates; act on Latin
   indices; AP is sure a  single field theory):
 \be\label{apsym}
  h^{*a}\!{}_\mu(y)=\kappa s^a{}_b h^b{}_\nu(x)\partial
x^\nu/\partial y^\mu; \ \
 \ \kappa>0, \ \ \ s^a{}_b\in O(1,D-1),
 \ \  s^a{}_b,\kappa=\mbox{const} .
 \ee
 The metric here is just the next quadratic form on the basis of the
 Minkowski metric:
 \[
g_{\mu \nu }= \eta_{ab}h^a{}_\mu h^b{}_\nu , \ \
 \eta_{ab},\eta^{ab}=\mbox{diag}(-1,1,\ldots,1)\,.
\]
  Field equations of such a great symmetry can be composed using
 a covariant notation, that is, the usual covariant differentiation
 with symmetric Levi-Civita connection, `;', and the fundamental
 tensor of this theory (which is a way simpler than the curvature
 tensor; coma `,' denotes the usual partial derivative, while
 square brackets denote anti-symmetrization of indexes):
 \be\label{lambda}
 \Lambda_{a\mu\nu} = h_{a\mu,\nu}- h_{a\nu,\mu} = 2h_{a[\mu;\nu]} \
 \ (g_{\mu\nu;\lambda} \equiv0); \
  \Lambda_{a[\mu\nu;\lambda]}\equiv 0\,  . \ee
 This tensor has three
irreducible parts, including vector and skew-symmetric tensor of
rank three (sometimes, in a covariant context, we omit in
contractions matrices $\eta^{ab},\eta_{ab}, g_{\mu\nu},g^{\mu\nu}$
--- there they can be restored unambiguously):
 \be\label{phis}
\Phi _a = \eta_{bc}\Lambda_{b\mu\nu}h_{c}{}^{\mu}h_{a}{}^{\nu}=
 \Lambda_{bba}, \ \ S_{abc}=3\Lambda_{[abc]}=
\Lambda_{abc}+ \Lambda_{bca} + \Lambda_{cab} \, .
 \ee

 The simplest derivative covariant in GR is the Riemann tensor
 which has three irreducible parts: Ricci scalar $R$,
 Ricci tensor $R_{\mu\nu}$,
 and Weyl tensor; only the last is
 responsible for gravitational waves
 (polarizations) as the others are `fixed' by the field
 equations of GR.

 The situation is different in
 Riemann-squared gravities with a Lagrangian
 composed of the three invariants
quadratic in the Riemannian curvature; however, due to different
reasons, these modified gravities are all unappropriate [as well
as $f(R)$-gravities, $f(R)\,{\neq}\, R$].

For example, $(a\,{+}\,bR\,{+}R^2\!/\!2)$-gravity  leads to an
incompatible  system  of equations (the trace part,
$\mathbf{E}_\mu{}^\mu\,{=}\,0$, can be used  that to remove the
term with $R_{;\l;\l}$ --- excepting the case $D\,{=}\,1$):
 \be\label{R2}
\mathbf{E}_{\mu\nu}=R_{;\mu;\nu} - R_{\mu\nu}(b+R)+g_{\mu\nu}
 (a/ 2 + bR/2+ R^2\!/4 -R_{;\l;\l})=0,
\ee
\[
\mathbf{E}^\star_{\mu\nu}=\mathbf{E}_{\mu\nu}
 + \varkappa\, g_{\mu\nu} \mathbf{E}_{\l}{}^{\l}=
 R_{;\mu;\nu} -
R_{\mu\nu}(b+R)+g_{\mu\nu}f^\star\!(R)=0.
\]
  The next combination of prolonged equations,
 $ \mathbf{E}^\star_{\mu\nu;\lambda} -
 \mathbf{E}^\star_{\mu\lambda;\nu } =0$,
after cancellation of the principal derivatives (5-th order),
gives
  new 3-d order equations which are irregular in the second jets:
 the term
 $ R_{;\ve}R^{\ve}{}_{\mu\nu\lambda} $
 can not be cancelled by the other terms which contain only
 the Ricci tensor
 and scalar.
The rank of these subsystem depends on the second derivatives,
 $g_{\mu\nu,\lambda\rho}$
(see \cite{pommaret} for the definition of PDE's regularity).

The most interesting case, $R_{\mu\nu} G^{\mu\nu}$-gravity (the
Ricci tensor is contracted with the Einstein tensor), gives the
following compatible system:
 \be \label{rg}  -\mathbf{D}_{\mu\nu}=
 G_{\mu\nu;\lambda}{}^{;\lambda}+
 G^{\epsilon \tau} (2R_{\epsilon
\mu\tau\nu } - \fr12g_{\mu\nu}R_{\epsilon \tau }) =0;
 \ \, \mathbf{D}_{\mu\nu;\lambda}g^{\nu\lambda}\equiv 0\,.
\ee
 In linear approximation, there are simple evolution
equations for the Ricci tensor and scalar:
 $\ \square R=0, \
\square R_{\mu\nu}=0$.  Using the Bianchi identity,
 $R_{\mu\nu[\lambda\epsilon;\tau]}\,{\equiv}\, 0$,
  its prolongation and contractions,
\[ R_{\mu\nu[\lambda\epsilon;\tau];\rho} g^{\tau\rho}\equiv0, \
 R_{\mu\nu[\lambda\epsilon;\tau]}g^{\mu\tau}\equiv0, \]
 we  write the evolution equation
 of Riemann  tensor
(we'll need just the linear approximation):
 \be \label{riem}
  R_{\mu\epsilon\nu\tau;\rho;\rho}=R_{\mu\nu;(\epsilon;\tau)}-
 R_{\mu\tau;(\nu;\epsilon)}+R_{\epsilon\tau;(\mu;\nu)}
 -R_{\epsilon\nu;(\mu;\tau)}+ ({\rm Riem.}^2).
 \ee
 This equation is more complex: it has a linear source term (in its
 RHS)
 composed from the Ricci tensor. As a result, in the general case when
 the Ricci-polarizations do not vanish, the polarizations
 relating to the Weyl tensor [and responsible for gravitational
 waves and tidal
 forces; their number is usual $D(D\,{-}\,3)/2$ plus one extra
 (`spin zero' or `trace') polarization] should grow linearly with
 time, \\[-11mm]

  \[ a(t) = (c_0 + c_1 t) e^{- i \omega t} , \]
  while the linear  approximation is valid.
  The total number of polarizations is $(D\,{-}\,1)^2$ --
 forth order equations have much more voluminous a Cauchy
 problem.
 There
 are no imaginary frequencies, no exponentially growing `eigen
 vectors' (exponential growing would contradict
 the correctness of the Cauchy
 problem), or polarizations -- all eigen values are just
 $w^2=k^2 $, but some of them are doubly degenerate, and some
 amplitudes should linearly grow.

This means that the regime of weak gravity is linearly unstable,
as well as the trivial solution itself
 (\emph{i.e.}, in this theory,  \emph{nothing}
 is not so \emph{real}).
  Hence,
this theory is physically irrelevant, because we are still living
in conditions of very weak gravity. [Note that in General
Relativity, when the Ricci tensor is expressed
 through  the energy-momentum tensor
 (which does not expand into plane waves --
 with the dispersion law of light in a
 vacuum),
   equation~(\ref{riem}) describes the process of
  gravitational wave generation.]

This linear instability  does not contradict the correctness of
 Cauchy problem; the compatibility theory (Kovalevskaya's theorem
and its generalization; see  Pommaret's book \cite{pommaret})
gives easy answers about the Cauchy problem, number of
polarizations, and so on (especially easy for analytical PDE
systems).


 The third  possible term (see, e.g., \cite{on}),
  can be written [using 5-minor
 of the metric,
minor of corank five; the corresponding Lagrangian, Gauss-Bonnet
or Lovelock term, can be written using 4-minor\,
 $[\mu\nu,\alpha\beta,\gamma\delta,\varepsilon\tau]
 \equiv \partial^4(-g)/(\partial g_{\mu\nu}
 \partial g_{\alpha\beta} \partial g_{\gamma\delta}
 \partial g_{\varepsilon\tau} )$] as
\[ \mathbf{D}_{\!\!(3)}^{\mu\nu}=
 (-g)^{-1}[\mu\nu,\alpha\beta,\gamma\delta,\varepsilon\tau,\rho\phi]
 R_{\alpha\gamma\varepsilon\rho}R_{\beta\delta\tau\phi}\,;
  \ \mathbf{D}_{\!\!(3)}^{\mu\nu}{}_{\!;\nu}\equiv0\,.  \]
 Covariant differentiation of the minor divided by $(-g)$ is
 identically zero (only metric $g^{\mu\nu}$ is there), while
 differentiation of either Riemann tensor leads to application of
 Bianchi identity (due to contraction with a highly skewsymmetric
  -- separately, in the `row indexes', and in the `column indexes' --
  tensor).
 This tensor identically vanishes   if $D\leq4$ (some
 people called this a `tricky identity', if I am not mistaken):
 $\,\mathbf{D}_{(3)}^{\mu\nu}\equiv 0$,\,
 because 5-minor is identically zero in low dimensions (you should
 cross out five rows and five columns, but that is impossible when
 the metric  is just a 4x4 matrix).

 In higher dimensions, if this tensor is the main term, the
 equations are irregular in second jets (and hence unappropriate).
  If this
 term is an addition (to a 4-th order Ricci-squared gravity or
 $R_{\mu\nu}G^{\mu\nu}$-gravity), it does not change the
 conclusion about the linear instability of the Weyl
 polarizations (or do not cure the irregularity of
 the Ricci-scalar-squared gravity).

As regards AP, the simplest compatible second order system
(non-Lagrangian, as it does not contain a term like $h_{a\mu}L$)
 \be \label{example}
 {\bf E}^*_{a\mu}=\L_{a\mu\nu;\nu}=0 \
 [\mbox{i.e. } (h\L_a{}^{\mu\nu})_{,\nu}=0, \
 h=\det h^a_{\ \mu} =\sqrt{-g}\,; \ \,
 {\bf E}^*_{a\mu;\mu}\equiv0\,] \ee
    looks, after linearization, like a  $D$-fold Maxwell's
 equation, see eq.~(\ref{lambda}),
 where infinitesimal diffeomorphisms serve as a set of
 gauge transformations; so, the number of polarisation
 degrees of freedom in this case
 (as well as for  other  AP equations
 with  similar  identities) is
  $D(D\,{-}\,2)$.
 \\[-1mm]

\noindent
 {\bf\large 3. Co- and contra-singularities
  in AP, and the unique field equation} \\[-2.5mm]

  \noindent
 AP is more appropriate as a
modified gravity, or just a good theory with topological charges
and quasi-charges (their phenomenology, at some conditions and to
the certain extent, can look
 like a quantum field theory) \cite{on,tc}.

  There is one unique equation of AP (non-Lagrangian, with a
unique $D$) which solutions are free of arising singularities.
 The formal integrability (compatibility) test
 \cite{pommaret} can be extended to the cases of degeneration of
either co-frame matrix, $h^a{}_\mu$ (co-singularities),  or
contra-variant frame (or frame density of some weight), serving as
 a local and covariant test for singularities. This test singles
out the next, unique equation (and $D\,{=}\,5$  \cite{on}; see
eq.\,(\ref{lambda});\,
 $h\,{=}\,\det
h^a_{\ \mu}\,{=}\,\sqrt{-g}$):
 \begin{equation} \label{ue}
 {\bf E}_{a\mu}=L_{a\mu\nu;\nu}- \fr13 (f_{a\mu}
 +L_{a\mu \nu }\Phi _{\nu })=0\, ;
\end{equation}
 here \qquad \qquad $ L_{a\mu \nu }=L_{a[\mu \nu]}=
\Lambda_{a\mu \nu }-S_{a\mu \nu }-\fr23 h_{a[\mu }\Phi_{\nu]},
 \ f_{\mu\nu}=2\Phi_{[\mu;\nu]}=
\Phi_{\mu,\nu}- \Phi_{\nu,\mu} $. \\[-1.5mm]

\noindent
 One should retain the identities (for further
details see \cite{on,tc}):
 \be\label{iden}
 \Lambda_{a[\mu\nu;\lambda]} \equiv 0\,,
  \ \  h_{a\l}\Lambda_{abc;\l}\equiv f_{cb}\,
  (= f_{\mu\nu}h_c^{\,\mu} h_b^{\,\nu}), \ f_{[\mu\nu;\l]}\equiv0
  .\ee
 Equation ${\bf E}_{a\mu;\mu}=0$ gives a
 {\em Maxwell-like\/} equation: $\ (f_{a\mu}
 +L_{a\mu \nu }\Phi _{\nu })_{;\mu}=0$,
\begin{equation}\label{max}{
 \mbox{ or \ }
 f_{\mu\nu;\nu}=(S_{\mu \nu\l }\Phi _{\l })_{;\nu} \ \
[= -\fr1 2 S_{\mu \nu\l }f_{\nu\l},
 \mbox{ see eq-n (\ref{skew}) below} ] \, .}
\end{equation}
In reality, eq-n (\ref{max}) follows from the symmetric part only,
because the skewsymmetric one gives an identity; note also that
the trace part
 becomes irregular   if $D\,{=}\,4$ (forbidden $D$;
 the principal derivatives vanish):
 \be\label{skew}
2{\bf E}_{[\nu\mu]}=S_{\mu\nu\l;\l}=0, \ {\bf
E}_{[\nu\mu];\nu}\equiv 0; \ee
 \be\label{trace} {\bf E}_{\mu\mu}={\bf
E}_{a\mu}h_b^\mu\eta^{ab} =\fr{4-D}3 \Phi_{\mu;\mu}
 -\fr12\L_{abc}^2+\fr13S_{abc}^2+\fr{D-1}9\Phi_a^2=0.
  \ee
System (\ref{ue}) remains compatible under adding $f_{\mu\nu}=0$,
see (\ref{max});  this is not the case for the other covariants,
$S, \Phi$, or the Riemann curvature; the last relates to tensor
$\L$ as usually:
 \[ R_{a\mu\nu\lambda}= 2h_{a\mu;[\nu;\lambda]}; \
h_{a\mu}h_{a\nu;\l}=\fr12 S_{\mu\nu\l}-\L_{\l\mu\nu}.\]
GR is a special case of AP. Using 3-minors
 (corank-3),
  $\,  [\mu \nu ,\varepsilon \tau ,\alpha \beta ] \equiv
 \partial^3 (-g)/(\partial g_{\mu \nu }
\partial g_{\varepsilon \tau }\partial g_{\alpha \beta }) ,\,$
 and their skew-symmetry features, one can write the vacuum GR
equation as follows:
 \be \label{gr}{
 2(-g)G^{\mu \nu }\,{=}\,[\mu \nu ,\varepsilon \tau
]_{,\ve\tau}\,{+} (g'^2\!)=
 [\mu \nu ,\varepsilon \tau , \alpha
\beta ](g_{\alpha \beta ,\varepsilon \tau }+ g^{\rho \phi }\Gamma
_{\rho ,\varepsilon \tau } \Gamma _{\phi ,\alpha \beta })=
 \fr12[\mu \nu ,\varepsilon \tau , \alpha
\beta ]R_{\alpha \varepsilon \tau  \beta} =0.}\ee
 Similarly,  all (but one)
 AP equations  can be reshaped in such a way
 that 2-minors of co-frame,
\[ \Bigl(\!\!%
\begin{array}{c}
  \mu \,\, \nu \\   a \,\, b\\
\end{array}%
\!\!\Bigr) =\frac{\partial^2 h}{\partial h^{a}_{\ \mu}
\partial h^{b}_{\ \nu}}
= 2!\, h h^{\,\ \mu}_{[a} h^{\,\nu}_{b]}, \ \mbox{i.e., \ } \
 [\mu_1 \nu_1 ,\ldots, \mu_k \nu_k  ]=\frac1{k!}
 \Bigl(\!\!%
\begin{array}{c}\mu_1\,\cdots\, \mu_k \\
 a_1\,\cdots \, a_k \\ \end{array}%
\!\!\Bigr)
 \Bigl(\!\!%
\begin{array}{c}\nu_1\,\cdots\,\nu_k\\
  a_1 \, \cdots \, a_k \\  \end{array}%
\!\!\Bigr),
 \] completely define the coefficients at the principal
derivatives.

 For example, the simple  equation (\ref{example})
 gives \cite{on}
 \[{ h^{2}{\bf E}^*_{a}{}^{\mu }=-gg^{\alpha
\mu}g^{\beta \nu} (h_{a\alpha ,\beta \nu }-h_{a\beta,\alpha  \nu }
)+\cdots 
 = h_{a\alpha ,\beta \nu }[\vspace{1mm}\alpha
\mu ,\beta \nu \vspace{-1mm}] +(h^\prime{}^{2})\ .}\]
 Like the determinant,
$k$-minors ($k\,{\leq}\, D$)
 are multi-linear expressions in
 elements of co-frame matrix, $h^a{}_\mu$, and some minors do not
vanish when rank$\,h^a{}_\mu\,{=}\,D{-}1$.

 For any AP equation [including
 eq-ns~(\ref{gr}) and (\ref{example})],
 with the \emph{unique exception}, eq.~(\ref{ue}),
  (where only the skew-symmetric
 part participates in the identity and can be written
 with 2- and 3-minors,
 while the symmetric part needs 1-minors which vanish too
 simultaneously when the co-frame matrix degenerates), the
  principal terms keep regularity  (and
 the symbol $G_2$ remains involutive \cite{on})
  if\,
 rank$\,h^a_{\ \mu}\,{=}\,D{-}1$.
  This observation is important and relevant to the problem of
singularities; it means seemingly that the unique equation
(\ref{ue}) does not suffer of  co-singularities in solutions of
general position.

The other case,  contra-singularities \cite{on}, relates to
degeneration of a contravariant frame density of some weight:
 \be
\label{dens} { H_a{}^\mu= h^{1/D_*} h_a{}^\mu;
 H=\det H^a{}_\mu, \
h_a{}^\mu= H^{1/(D-D_*)} H_a{}^\mu\, .} \ee
 Here $D_*$ depends on the choice of
equation: $D_*=2$ for GR, $D_*=\infty$ for eq-n~(\ref{example}),
and $D_*=4$ for the unique equation (which can be written
3-linearly
in $H_a{}^\mu$ and its derivatives \cite{on}).  %

If integer, $D_*$ is the forbidden spacetime dimension.
 For the unique equation, the
nearest possible $D$,  $D=5$, is of special interest: in this case
minor $H^{-1} H^a{}_\mu$ simply coincides with $h^a{}_\mu$; that
is, a contra-singularity simultaneously implies a co-singularity
(of high corank), but that is impossible! The possible
interpretation of this observation is:
 for the unique equation,
 contra-singularities are impossible if $D\,{=}\,5$
 (perhaps due to some specifics of
\emph{Diff}-orbits on the $H_a{}^\mu$-space).
  This leaves no room for any changes in the theory (if
 nature abhors singularities).\\[-1mm]

\noindent
 {\bf\large 4. Stress-energy tensor and new gravity with a
 `weak Lagrangian';\\  {\em dwarf}, {\em normal},
 and {\em giant} (unstable) polarization degrees in AP} \\[-2.5mm]

  \noindent
  One can rearrange ${\bf E}_{(\mu\nu)}{=}\,0$ picking out
 (into  LHS) the Einstein tensor,
 but the rest terms are not
 a proper stress-energy tensor: they contain linear terms
 $\Phi_{(\mu;\nu)}$ [no positive energy (!)]:
 \be\label{sym} {\bf E}_{(\mu\nu)}+ 2g_{\mu\nu}{\bf E}_{\l\l}=
 -G_{\mu\nu}-\fr23\Phi_{(\mu;\nu)}+(\L^2\mbox{-terms})=0.
  \ee
 However, the prolonged equation
${\bf E}_{(\mu\nu);\l;\l}$ can be written as
$R_{\mu\nu}G^{\mu\nu}$-gravity (\ref{rg}):
 \be \label{tmunu}{
 G_{\mu \nu
;\lambda ;\lambda }+ G_{\epsilon \tau} (2R_{\epsilon \mu \tau \nu
} - \fr12g_{\mu \nu }R_{\epsilon \tau }) =T_{\mu\nu} (\Lambda
'^{2},\cdots), \ T_{\mu\nu;\nu}=0; }\ee
 up to
quadratic terms,  \ ${ T_{\mu\nu}\,{=}\,
\fr29(\fr14g_{\mu\nu}f^2\,{-}\,f_{\mu\l}f_{\nu\l})\,
 {+}\,B_{\mu\eps\nu\tau}(\L^2){}_{,\eps\tau}
 } $\, \cite{on};
tensor $B$ has symmetries of the Riemann tensor, so  term $B''$
adds nothing to the $D$-momentum and angular momentum.

This equation (\ref{tmunu}) follows also from the least action
principle. The `weak Lagrangian' (the term of N.Kh.\,Ibragimov for
the case when variation is zero due to both  field equations and
their prolongations) is quadratic in the field equations,
 {\em i.e.}
is trivial\footnote{ This triviality, however, is of another sort
than the triviality of surface terms.}
 [one should use the trace eq-n (\ref{trace}), and the
identity  $R\,{=}\,{-}2\Phi_{\mu;\mu}{+}\, (\L^2)$; $D\,{=}\,5$]:
  \be \label{lagt}L=  {\bf E}_{(\mu\nu)}^2 -7{\bf E}_{\l\l}^2
  \equiv  R_{\mu\nu}G^{\mu\nu} + \fr19 f_{\mu\nu}^2 +
 \fr49[(3G_{\mu\nu}-\Phi_{\mu;\nu})
  \Phi_{\mu}+\Phi_{\l;\l}\Phi_\nu]_{;\nu}+
 (\L'\L^2,\, \L^4). \ee
 The main, quadratic
terms, after exclusion of covariant divergences (surface terms),
look like a modified gravity (higher terms can add to $T_{\mu\nu}$
only a trivial quadratic contribution, like $B''$.

 This Lagrangian is trivial, as well as all its Noether currents;
  this also means that the  contribution of gravitation
  to the `total energy' is
  negative and the `total energy' is strictly zero.
  (All this Lagrangian issue follows as a mere bonus, without any
  {\em ad hoc\/} activity\,!)

 Note that only $f$-covariant (three transverse polarizations in
$5D$) carries $D$-momentum and angular momentum
 ({\em ponderable} or {\em tangible} waves);
  other 12 polarizations are {\em imponderable}, or
 {\em intangible}. This is a very strange thing (it is scarcely
 possible in the Lagrangian tradition).

 These $f$-waves feels only the metric and $S$-field,
 see (\ref{max}), but $S$
has effect only on polarization (`spin') of these waves:
$S_{[\mu\nu\l]}$ does not enter the eikonal equation, and
$f$-waves moves along the usual Riemannian geodesics.

 However, $f$-component is not the usual
(quantum) EM-field, it's just an important covariant responsible
for energy-momentum (there is no gradient invariance for $f$
\cite{on}).

 Another important feature is the linear instability
 of the trivial solution:
 some {\em intangible} polarizations grow linearly with time
 in the presence of
 {\em tangible} $f$-waves. Really,  the linearized
 eq-n~(\ref{ue}) and identity (\ref{iden}) yield
 (the following equations should be understood as linearized):
  \be \label{inst}
 3\Lambda_{abd,d}= \Phi_{a,b}-2\Phi_{b,a} \
 (\mbox{trace part:} \, \Phi_{a,a}\,{=}\,0),
 \ \Lambda_{a[bc,d],d}\equiv0, \ \Rightarrow \
 \Lambda_{abc,dd}=-\fr23 f_{bc,a}\, . \ee
 The last D`Alembert equation has a \emph{source} in its RHS.
  Some components of $\Lambda$
 (most symmetrical irreducible parts,
 as well as the Riemann curvature)
 do not grow because
 (linearized equations again)
 \[{ S_{abc,dd}=0, \ \Phi_{a,dd}=0, \ \,
   f_{ab,dd}=0, \ R_{abcd,ee}=0. }\]
  However the least symmetrical $\L$-components
  (triangle Young diagram), in fact only three polarizations
  of them
  which are to be called $\L^{\!\bullet}$-waves
  (three growing but intangible
   polarizations), do go up with time
  if  the ponderable waves (three $f$-polarizations)
  do not vanish. This should be the case for
  solutions of general position.
  These {\em giant\/} polarizations,
  $\L^{\!\bullet}$-waves,
  should result in strong nonlinear effects, and it is of special
  interest if some space regions can witness
  more $f$-waves and hence more
  instability,
  more nonlinearities, in comparison with other regions.

  The forth polarization of vector $\Phi_\mu$ [the fifth
  one is eliminated by the trace equation, (\ref{trace})] is the
  (only) longitudinal polarization;
  it  relates to the gradient
  part: $\Phi_{\mu}\,{=}\,\Psi_{,\mu}$. One can formally write the
  evolution equation for the longitudinal polarization
   [see eq-n~(\ref{trace})],
  \ $ \square \Psi = \L^{\!\bullet}{}^2+\cdots\, $;

  \noindent
  so, the giant polarizations squared
  do influence the longitudinal polarization.

  [Interestingly, the linearized equations (\ref{ue}) loose its
  trace part if $D\,{=}\,4$ (forbidden dimension; still one can add
  eq-n $\Phi_{\mu,\mu}\,{=}\,0$ `by hand') and in this case there
  is a new symmetry --- with respect to infinitesimal conform
  transformations which serve as a kind of
   gradient transformations of
  vector $\Phi_\mu$, and, therefore, eliminate the longitudinal
  polarization, so to say.]

  The skew-symmetric tensor $S$ is responsible for three
  polarizations. One can introduce  pseudo-tensor
  (remember $D\,{=}\,5$)
   \[ \tilde{f}_{ab}=\fr16\,\ve_{abcde}S_{cde}; \]
 then, from eq-n (\ref{skew}) and the totally skew-symmetric part
 of identity (\ref{iden}), it follows (again a Maxwell-like system):
 \[ \tilde{f}_{[\mu\nu;\lambda]}=0\,, \ \,
   \tilde{f}^{\mu\nu}{}_{;\nu}=\fr18\,h^{-1}\,
   \ve^{\mu\nu\l\ve\tau}\L_{a\nu\l}\L_{a\ve\tau}\,.\]
 So, we have just three $S$-polarizations.

 Three $\L^{\!\bullet}$-polarizations correspond to $\L$-tensor of
 a specific, gradient (or rotor-gradient) form:
 \(\, \L_{\ve\mu\nu}\,{=}\,A_{[\mu,\nu];\ve}\,.\)
 At last, there remain five polarizations; this is just  the
 number of ordinary gravitational (Weyl) polarizations (in 5$D$);
 the evolution of
 these waves, see eq-ns (\ref{riem}) and (\ref{sym}), again has
 $\L^{\!\bullet}{}^2$-terms in its RHS (as a source) ---
 this time organized
 as a tensor relating to the square Young diagram (symmetry of the
 Weyl tensor).

 The current for the $f$-waves is just $Sf$-term, see (\ref{max}),
 therefore these waves are most weak, {\em dwarf},
  that is, their amplitude,
 $a_f$,
 should be smaller than the amplitudes of all other polarizations,
  \ $ a_{\L^{\!\bullet}}\gg a_{W}, a_S, a_L \gg a_f\, \
 \, (3_{\L^{\!\bullet}}+5_W+3_S+1_L+3_f=15) $.

 If some form of reduction to a 4$D$ picture takes place, there
 could come forth  eight `preferable for 4$D$' (or not so sensible to
 the extra dimension) polarizations:
 $2_{\L^{\!\bullet}}+2_W+1_S+1_L+2_f=8$.
 \\[-1mm]

\noindent
 {\bf\large 5.
  Expanding O$_4$-wave and cosmology;
  topological (quasi-)charges} \\[-2.5mm]

  \noindent
    The great symmetry of AP equations gives scope for symmetrical
 solutions. In contrast to GR,  eq-n~(\ref{ue}) has
non-stationary spherically
 symmetric solutions (as an example of longitudinal waves).
An $O_4$-symmetric field  can be generally written
 \cite{on} as\\[-0.4mm]
\begin{equation}  \label{spsy}{
 h^{a}{}_{\mu }(t,x^i)=
  \left(%
\begin{array}{cc}
  a & b\,n_i \\
  c\,n_i & e\,n_i n_j+ d\Delta _{ij}\\
\end{array}%
\right); \  \ n_i=\frac{x^i}{r}; }
\end{equation}
 here $i,j\,{=}\,(1,2,3,4)$,
  $a,\ldots,e$ are functions of time, $t=x^0$, and radius
 $r$, $\Delta_{ij}\,{=}\,\delta_{ij}\,{-}\,n_i n_j, \
 r^2\,{=}\,x^i x^i$.

 As functions of radius, $b,c$ are odd
  while the others even;
  the boundary conditions are: $e\,{=}\,d$ at $r\,{=}\,0$,
 and $h^a{}_\mu\,{\to}\, \delta^{\,a}_\mu$ as
  $r\,{\to}\, \infty$.
Placing in (\ref{spsy}) $b\,{=}\,0, e\,{=}\,d$ (another
interesting choice is $b\,{=}\,c\,{=}\,0$)
 and making integrations, one arrives to the next system
 (it resembles Chaplygin gas dynamics; dot and prime
 denote  time and radius derivatives, respectively.)
\begin{equation}\label{gas}
\dot{A}=AB' -BA' +\fr{\,3}r AB,
 \  \dot{B} =AA' -BB' -\fr{\,2}r B^{2},
\end{equation}
where $A\,{=}\,
 a/e\,{=}\,e^{1/2},\ B\,{=}\,{-}\, c/e\, $.
This system has non-stationary solutions, and a single-wave
solution (of {\em proper sign}) might serve as a suitable (stable)
cosmological expanding background. 
  The condition $f_{\mu\nu}\,{=}\,0$ is a must for
   solutions with such
 a high symmetry (as well as
 $S_{\mu\nu\l}\,{=}\,0$); so, these $O_4$-solutions
 carry no energy, weight nothing ---
 some lack of \emph{gravity}\,!

 A more realistic cosmological model might look like a single
 $O_4$-wave
 (or a sequence of such waves) moving along the radius and being
 filled with a sort of chaos, or an ensemble of chaotic waves,
 both tangible (\emph{dwarf}; $a_f\,{\ll}\,1$) and
  intangible ($a_{\L^{\!\bullet}}\,{<}\,1$,
  but intense enough that  to give non-linear
  fluctuations with $\delta h\,{\sim}\,1$).
  Development and examination of stability
 of this model is an interesting problem.
 The metric inhomogeneity in such a cosmological $O_4$-wave
 can serve as a slowly varying
  {\em shallow dielectric waveguide} for that dwarf
 $f$-waves \cite{on}. The ponderable  waves
  should have wave-vectors almost tangent to
the $S^3$-sphere of the wave-front that to be trapped inside this
spherical shell; the {\em giant} waves can grow up, and partly
escape from the waveguide,
 and their wave-vectors
can be some less tangent to the $S^3$-sphere.
The shell thickness can be small for an observer in the center of
$O_4$-symmetry, but in co-moving coordinates it can be very large,
but still much smaller than the current radius of the spherical
shell,  $L\,{\ll}\,R$.

 This picture leads to the anti-Milne cosmological model,
  $a\,{=}\,a_0(1\,{+}\,H_0\,t),   \ k\,{=}\,{+}1$,
  with the next simple equation of distance modulus
  (it's good for SNe Ia/GRB
  data;\footnote{See some diagrams
 in zhogin.narod.ru/pirt11.pps, or \cite{tc}b.}
  $d_*=10$\,pc):
 \[ \mu(z)=\mu_0+5\log[(1+z)\,\ln(1+z)], \
 \mu_0=-5\log( H_0d_*/c)\approx 43.3\,. \]

 This model does well the {\em job of inflation}.
  Only very small part
 of the spherical shell corresponds to
 $z_{\rm CMB}\sim10^3$ (decoupling of CMB):
 $ \varphi \simeq\frac1{\G }\, \ln(1+z) \
 (\g\mbox{-factor} \
\G\gg1 ),\ \ \varphi_{\rm CMB}\,{\ll}\, 1$.
 Moreover, very
 separated, even  opposite  points of the shell
 (at such $z$) are not
 causally independent -- they have the common past
 along the extra
 dimension. 

 The symmetry of this cosmological background is very high,
 enabling an interesting set of topological quasi-charges
  [localised field configurations of some (sub)symmetry, carrying a
 discrete feature --- a topological quasi-charge], and some
 phenomenology of {\em topological quanta\/} on expanding, chaotic
 background should emerge.  Because of
 time/volume limitations I will not settle in detail this subject
 (just see \cite{on}, \cite{tc}a).
 Still we should correlate somehow the `true' (or naturally
 geometrical) tensor $T_{\mu\nu}$ of eq.\,(\ref{tmunu})
 (i.e.\ its {\em quantum\/} part  which arises while
   topological quanta scatter
 and disturb the chaotic ensemble of perceptible $f$-waves) with
 the `phenomenological energy-momentum' of GR.
 In units $\hbar\,{=}\,1, c\,{=}\,1$,
  the `phenomenological momentum'
 of a particle is just its wave-vector, but the `true' momentum,
 sure being proportional to the wave-vector of quantum's
 psi-function,\footnote{ The superposition principle emerges due to
 (a) the huge size $L$ of quanta along the extra-dimension and (b)
 the fact that $f$-waves are almost tangential in the shell
  [so, some scattering amplitudes (framing vectors or something
  similar) of different parts along the extra dimension,
  with the same projection (i.e.\ cophased), should be summed up].}
  should include the small factor $a_f^2$, which defines
  the overall scale of the perceivable, true-energy carrying waves.
 So, a rude estimation is possible:
  \be\label{planck}
 T_{\mu\nu} \approx a_f^2 T_{\mu\nu}^{\rm (phen.)},
 \ \mbox{ and (see the next section) } \ \l_{\rm Planck}^2\approx a_f^2 L^2\, .
 \ee
 It seems that, in this theory,
 the Planck length is not of a fundamental sense
  (and the spectrum of chaotic waves should not
 continue to such a small wavelength).
 \\[-1mm]

\noindent
 \textbf{\bf\large 6. Newton's gravity changes;
 suppression of short gravitational waves
  } \\[-2.5mm]

  \noindent
  A massive body in this theory (assuming that it is right) should
 be of great length along the extra dimension, and we would like to
 estimate the behavior of gravitational potential
 (the case of weak static field), and possible
 deviation from  Newton's law of gravity.

  Let us start with a point mass; the  `new gravity',
   eq.\,(\ref{tmunu}), gives a
   4d (from 5D) bi-Laplace equation with
a $\d$-source,
 and its solution ($R$ is 4d distance, radius) is easy to find:
\be\label{point} {
 \triangle_{(4)}^2\varphi = -\,\frac a {R^3}\,\d(R); \ \,
\varphi(R^2) = \frac a{\,8\,}\ln R^2 -\frac b {R^2} \ (+\,c\,,
\mbox{ but $c$ does not matter}); } \ee
 the  force
between two point masses is $ F_{\rm point} = \frac a{4R}
+\frac{2b}{R^3}$ ($a,b$ are proportional to  both masses).

Now let us suppose that all masses are distributed along the extra
 dimension with a {\em universal function},
  $\mu(p), \ \int\! \m(p)\,d p =1$.
  Then the attracting (gravitational) force takes the next form
[see (\ref{point}); $r$ is usual 3d distance; note that $V(r)$ can
be restored if $F(r)$ is measured]:
 \be \label{mon}
F(r)= \frac{d}{dr}
\! \int\! \!\!\int^{{}^\infty}_{\!\!\!{}_{- \infty}} \!\!\!\!\!
\varphi(r^2+(p-q)^2)\,\mu(p)\,\mu(q)\,dp\,d q =\frac{a\,r}4\, V-
b\, V\,' , \ \, V(r)\,
{=}\int\!\!\!\int\!\frac{\mu(p)\,\mu(q)\,dp\,d q}{r^2+(p-q)^2}.
 \ee
Taking $\m_1(p)=\pi^{-1}/(1+p^2)$ (typical scale along the extra
dimension is taken as unit, $L=1$), one can find $rV_1(r)=1/(2+r)$
and (note, if $ a\,{=}\,b$, the Rindler term, $\sim r^0$,
vanishes)
 \[
F(r)=\frac{a}{8+4r}+\frac{2b(1+r)}{r^2(2+r)^2}
 =\frac b {2r^2} +
\frac{a(2+r)-2b}{4(2+r)^2}.\]
~\\[-22mm]

 \noindent    
\begin{figure}[h]
 ~~\includegraphics[bb=25 10 290 230,height=65mm]{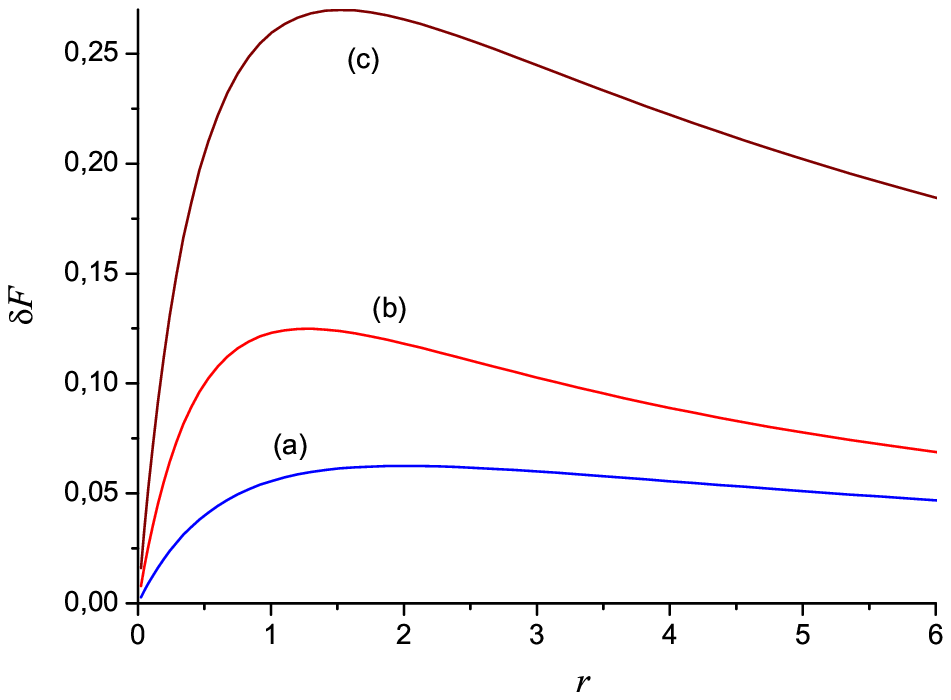}
  \end{figure}\\[-62mm]
\noindent
\begin{tabular*}{170mm}{p{78.5mm}p{85.5mm}}
  &  
  {\small {\bf Fig.\,1}: curve\,(a) shows $\d F= F{-}1/r^2$,
see  eq.~(\ref{mon})  and text below [$a{=}b\,({=}2) $
 is chosen to ensure $\d F (0){=}0$;
more exactly, this is the dimensionless deviation from Newton's
 law,
 $(F/F_N -1)L^2/r^2$
as a function of $r/L$];

\noindent curves (b), (c) correspond to
$\mu_2{=}2\pi^{-1}\!/(1{+}p^2)^2$,
$\mu_3{=}2\pi^{-1}p^2/(1{+}p^2)^2$;
 $a/b$ is chosen also that to ensure
$\d F (0){=}0$.

\vspace{2mm} We see that, in principle,
 this theory can explain
galaxy rotation curves, $v^2(r)\,{\sim}\, rF
\stackrel{r\to\infty}{\longrightarrow}\,$const,
 without need for Dark Matter or MOND. }
 \vspace{5mm}
\end{tabular*}
\vspace{1mm}

 However, at this  stage it is not easy to give an estimate for
 the length scale $L$: it depends on the mass distribution both
 along the extra dimension and in the ordinary space.
  Note, for example, that
 the usual feature of Newton's law of gravity,
  that a spherical massive shell
 has no effect on inner massive bodies, is no more true.

 Generation of  the most representative components of Riemann
 tensor (i.e., gravitational waves, GW) is described,
 a bit schematically, by the next
  equation [see eq-s\,(\ref{riem}),\,(\ref{tmunu});
 $\alpha,\beta$ -- space indexes; full space derivatives,
 $()'$,
 do not matter in the RHS]:
 \be \label{riem2}
 \square R_{0\alpha0\beta}\simeq \ddot{R}_{\alpha\beta}+()';
 \ \ \ddot{R}_{\alpha\beta}
 \sim T_{\alpha\beta} +
 \ddot{B}_{0\alpha0\beta}(\L^{\!\bullet\,2})+()',
 \ \mbox{ while in GR: }
 \, \sim \l_{\rm Planck}^2
 \ddot{T}^{\rm (phen.)}_{\alpha\beta} +()'.
 \ee
 So, one can suggest that, while
$L^2 \ddot{T}_{\alpha\beta}\,{\gg}\,T_{\alpha\beta}$
 (that is, for
short waves, $\l\,{\ll}\,L$),
 generation of gravitational waves, by
virtue of $T_{\mu\nu}$,
 in the new gravity is much `weaker' than in GR.

 However, it seems that the giant polarizations can
  also contribute to the process (GW generation). They
  should form  a kind of halo, a disturbance of size $L$,
 near a heavy body; the form of this halo is either cusped or
 cored -- depending on presence or absence of the Rindler term.

 The  00-component of the symmetric part
  (\ref{sym})
 gives an equation for the gravitational potential, as follows
 (neglecting differentiation along the extra dimension, near the
 middle of the shell; the static problem):
  \[  
 \triangle_{(3)}\varphi \sim <\L^{\!\bullet\,2}>_{00}\, .
 \]  
  Near a body, the non-Newtonian part of potential  behaves like
  \[\mbox{ either } \
  \d \varphi \sim r, \ \mbox{ hence }
  <\L^{\!\bullet\,2}>_{00}\sim 1/r \mbox{ (cusped)}, \]
  \[ \mbox{or \ \ \ \ }
  \d \varphi \sim r^2, \ \mbox{ hence }
  <\L^{\!\bullet\,2}>_{00}\sim  \mbox{const \  (cored)}; \]
  in any case, at large scales, $r\,{>}\,L$,
   the halo drops:
  $\d \varphi \sim \ln r, \ \mbox{ hence }
  <\L^{\!\bullet\,2}>_{00}\sim 1/r^2$.

  The absence of divergency (no cusp, but core)
   seems a natural requirement,\footnote{There is a subtlety here;
   the force includes two integrations along the extra-dimension,
   while the potential only one (so, some small Rindler term
   still may
   survive). } and in this case the generation of short GW, by
   virtue of such a halo, is also suppressed (exponentially).

   This difference between these two  gravities, with
   respect to generation of GW, can in principle be tested: the
   method based on pulsar timing, that
    to observe very long (nHz) gravitational waves,
    is actively discussing (see e.g.\ \cite{pulsar}).

\end{document}